\documentstyle[twocolumn,prb,epsf,aps]{revtex}
\begin{document}
\title{Magnons and Skyrmions in Fractional Hall Ferromagnets}
\author{A.~H.~MacDonald$^1$ and J.~J.~Palacios$^{1,2}$}
\address{$^1$Department of Physics, Indiana University, 
Bloomington, IN 47405, USA}
\address{$^2$Departamento de F\'{\i}sica Te\'orica de la Materia Condensada,
Universidad Aut\'onoma de Madrid, Cantoblanco, Madrid 28049, Spain}
\date{\today}
\maketitle

\begin{abstract}

Recent experiments have established a qualitative difference
between the magnetization temperature-dependences $M(T)$ of 
quantum Hall ferromagnets at integer and fractional
filling factors.  We explain this difference in terms 
of the relative energies of collective magnon and 
particle-hole excitations in the two cases.
Analytic calculations for hard-core model systems are 
used to demonstrate that, in the fractional case, interactions 
{\it suppress} the magnetization at finite temperatures 
and that particle-hole excitations rather than long-wavelength
magnons control $M(T)$ at low $T$. 

\vspace{0.3cm}
\end{abstract}

The many-particle physics of two-dimensional (2D) electron systems
in a strong perpendicular magnetic field is enriched by 
the macroscopic degeneracy of a Landau band. 
Interactions in this system cannot be treated perturbatively and 
the electronic system can occur in one of a variety\cite{dassarmapinczuk}
of non-Fermi-liquid states, depending on the Landau level
filling factor $\nu = N/N_{\phi}$.  (Here $N$ is the number
of electrons, $N_{\phi} = A B / \Phi_0 $ is the degeneracy
of the Landau band, $A$ is the system area, $B$ is the magnetic
field strength, and $\Phi_0$ is the electron magnetic flux quantum.)
Our interest here is in the strong ferromagnet ($S=N/2$) states which occur 
at $\nu = 1/m$ and $\nu = 2 -1/m$ for odd integers $m$.\cite{smgahminbook}
Interest in these states has grown with the development of 
NMR\cite{BarrettNMR} and other\cite{other} techniques which allow 
the spin-magnetization $M(T)$ and other spin-dependent properties 
to be measured.  Until recently, experimental data on $M(T)$ has been 
available only at $\nu = 1$ state.  In this case $M(T)$ remains
large at temperatures $T$
well above the Zeeman temperature $T_z = g^{*} \mu_B B/ k_B 
\equiv \Delta_z/k_B $.
(Here $\mu_B$ is the electron Bohr magneton and $g^{*}$ is the 
host semiconductor g-factor.)  The present work is motivated by
new experimental results\cite{barrettyale} which show that for $\nu =1/3$, 
both the shape of the $M(T)$ curve and the temperature scale on which 
$M$ drops to small values are altered.   The objective of this
work is to explain this difference on a qualitative level. 
We propose that the difference is due to inversion of the relationship between
collective magnon and particle-hole excitation energies and 
contrast the microscopic physics of quantum Hall ferromagnets
at integer and fractional filling factors, by discussing the 
properties of hard-core model systems.
Analytic calculations for these models are  
used to demonstrate that, in the fractional case, interactions
{\it suppress} the magnetization at finite temperatures.
Exact diagonalization calculations for realistic interaction models support the 
applicability of hard-core model qualitative considerations to
experimental systems. 

In the quantum Hall regime, the kinetic energy term in the
Hamiltonian enters only through the constraint that the 
interacting electrons must lie in the lowest available 
orbital Landau level.  This constraint is most conveniently 
enforced in the symmetric gauge\cite{leshouches} for which the allowed 
single-particle orbitals are eigenstates of $L_z$ and have the form 
\begin{equation} 
\phi_m(z) = \frac{z^m}{(2^{m+1} \pi m!)^{1/2}} \exp ( - |z|^2/4)
\label{singpart}
\end{equation}
where $z = x - i y$ is the complex 2D coordinate,
$m = 0,1, \ldots , N_{\phi-1}$ and 
we have chosen $\ell \equiv (\hbar c /e B)^{1/2}$ as the 
unit of length.  Pairs of electrons also have only one
available relative motion state for each angular momentum,
$\phi_m^{rel}(z-z') = \phi_m((z-z')/\sqrt{2})/\sqrt{2}$. 
Consequently the interaction term in the Hamiltonian is 
completely characterized by the 
Haldane pseudopotentials\cite{reftohaldane}
$V_m \equiv \langle \phi_m^{rel} | V(r) | \phi_m^{rel} \rangle$. 
For realistic interaction models, $V_m$ decreases monotonically
with $m$.  Interaction physics in the quantum Hall regime 
is often usefully addressed by considering hard-core models where only
the lowest few $V_m$ are non-zero, and $V_{m+1}/V_{m}$ 
is small enough that each successive pseudopotential defines  
a separate energy scale and resolves the many-particle 
energy spectrum more finely.  

A useful orientation is the hard-core model with $V_0$ and 
$V_1$ arbitrarily large and all other pseudopoentials set to zero.
For this model, finite energy eigenstates necessarily have zero
probability amplitude for finding pairs of electrons in 
states with relative angular momenta $0$ and $1$, and 
therefore can be written in the form\cite{murray,dziarmaga}
\begin{equation}
\Psi[z] = \big[ \prod_{i<j}^{N} (z_i - z_j)^2 \big] \Psi^{*} [z]
\label{v0v1wf}
\end{equation} 
where $\Psi^{*}[z]$ is a many-fermion spinor.  Since the 
maximum single-particle $m$ at $\nu = 1/3$ is 
$N_{\phi} = 3 N$, and the factor in square brackets in 
Eq.(~\ref{v0v1wf}) contains powers of $z_i$ up to $2 N$ it 
follows that the many-fermion spinors $\Psi^{*}[z]$ can be 
composed of orbitals with $m$ from $0$ to $N$.   
This establishes a one-to-one mapping between the 
zero energy eigenstates of the $V_0 - V_1$ hard-core
model filling factor  $\nu = \nu^* +2 $ and the full many-fermion Hilbert
space at $\nu =\nu^*$.  For $k_B T$ much smaller than
$V_0$ and $V_1$, the hard-core model $M(T)$ at $\nu = 3$
will be identical to $M(T)$ for a 
non-interacting electron system at $\nu = 1$:
\begin{equation}
M^{HC}(T) = N \mu_B \tanh (\Delta_z / 4 k_B T).
\label{v0v1moft}
\end{equation}
This result should be compared with that non-interacting electrons
at the same filling factor:   
\begin{equation}
M^{NI}(T) = N \mu_B \frac{ \sinh (\Delta_z / 2 k_B T)}
{ z + \cosh(\Delta_z / 2 k_B T)}.
\label{nonintmoft}
\end{equation} 
where $z = \exp(\mu/k_B T )$ is the fugacity which, for $\nu =1/m$,
approaches its maximum value  $1/(2m-1)$, as $T \to \infty$:
\begin{equation}
\frac{1}{m} = \frac{z}{\exp(\Delta/2 k_B T) +z} +
\frac{z}{\exp(-\Delta/2 k_B T) +z}
\label{fugacity}
\end{equation}
It follows that hard-core model interactions {\it always} reduce $M(T)$,
at $\nu = 1/3$, in conflict with the common qualitative notion of exchange
and correlation enhanced spin magnetization.  

We now consider an extended hard-core model for which 
$V_0$ and $V_1$ are arbitrarily large, 
but $V_2$ is also non-zero.  We will argue later that this model 
captures the essence of the physically realistic situation.
The added pseudopotential lifts the degeneracy among
the low energy eigenstates of the hard-core model.  In Table I
we list many-particle eigenenergies for the ground state 
at $\nu =1/3$ and for states with fractionally charged 
quasielectrons and quasiholes and various integer numbers
(K) of reversed spins.  
We first note the gross differences between the quasiparticle
and quasihole results.  Zero interaction energy quasihole
states occur at all values of $K$ and are related to the zero 
energy states of the $\nu = 1$ $V_0 \ne 0$ hard-core model 
by the same $\prod_{i<j} (z_i - z_j)^2$ factor that 
appears in Eq.(~\ref{v0v1wf}).  In the $\nu=1$ case, the large 
$K$ zero energy quasihole states correspond\cite{nu1hardcore} to 
the skyrmion\cite{sondhi} topologically charged spin-textures
of classical field theories.  We follow previous practice and refer to
all the quasiholes with $K \ge 1$ as skyrmion states.  For 
$\nu = 1$ particle-hole symmetry\cite{kyphsym} guarantees 
that skyrmion quasiparticle states appear for $K \ge 2$
and that these have the same dependence of energy on $K$ 
as the quasihole states.  
We can infer from Table I that, although quasiparticle-quasihole
symmetry does not hold for $\nu = 1/3$, quasielectron skyrmion
states do occur and that, just as at $\nu =1$, those with 
larger numbers of reversed spins have lower interaction energies.
The $K=0$ quasielectron state at $\nu = 1/3$, which is pushed to 
infinite energy in 
both hard-core models, is analogous to higher Landau level 
quasiparticles at $\nu=1$. 
The lack of any dependence of the quasihole interaction 
energy on $K$, means that the lowest energy quasiholes will 
have $K=0$ at any finite Zeeman coupling strength for this model.  

The excitation spectrum at $\nu=1/3$ will include both 
particle-hole excitations and spin-wave collective excitations 
whose energy is given at long-wavelengths by\cite{GM95} 
\begin{equation}
\epsilon_{SW}(q) = \Delta_z + 4 \pi \rho_s q^2 /\nu  
\label{swdisp}
\end{equation}
where $\rho_s$ is the spin stiffness and $\Delta_z = 
g^{*} \mu_B B $ is the Zeeman coupling gap. $\rho_s$ also determines
the energy of large-$K$ skyrmionic particle-hole excitations:\cite{sondhi,GM95}
\begin{equation} 
\Delta_{SK} = \Delta_z (K_{e} + K_{h}) + 8 \pi \rho_s 
\label{skgap}
\end{equation}
where $K_{e}$ and $K_{h}$ are the number of reversed spins 
in the quasielectron and quasihole skyrmions respectively.
Insight into the physics which determines $\rho_s$ for a given
electron-electron interaction can be gained by considering the 
single-mode approximation expression\cite{rasolt,mitra} for 
$\epsilon_{SW}(q)$:
\begin{equation}
\epsilon_{SW}(q) = \Delta_z + 
4 \sum_m g_m \big[ \sum_k |T_{k,m}(q)|^2  V_k - V_m \big].
\label{swsma}
\end{equation}
Here $g_m$ is the probability\cite{mitra} in the ground state
for a pair of electrons to be in a state of relative angular momentum $m$;
normalized such that the pair distribution functions is 
\begin{equation}
g(r) = \frac{8 \pi \ell^2}{\nu}  \sum_m g_m |\phi_m^{rel}(\vec r)|^2 
\label{gofr}
\end{equation}
where $\phi_m^{rel}(\vec r)$ is a relative motion wavefunction.\cite{leshouches}
$T_{k,m}$ is the coefficient of $\phi_k^{rel}(\vec r)$ in the 
expansion of  $\phi_m^{rel}(\vec r + \vec q \ell^2 \times \hat z)$:
\begin{equation}
|T_{k,m}(q)|^2 =  \exp( - q^2 \ell^2 /4) \frac{k!}{m!} 
[L_k^{m-k}(q^2 \ell^2 /4)]^2 (q^2 \ell^2 /4)^{m-k}.
\label{translation}
\end{equation}
Eq.(~\ref{swsma}) shows that, just as at $\nu =1$\cite{BIE81,KH84},
spin-wave excitations change the relative distance between
pairs of electrons and thereby increases the interaction 
energy (4 $\sum_m g_m V_m$) from its minimum ground-state
value.  For the case of the extended hard-core model the 
ground state at $\nu = 1/3$ is the Laughlin wavefunction\cite{laughlin}
for which $g_0 = g_1 = g_2 =0$.  Comparing Eq.(~\ref{swsma}) and 
Eq.(~\ref{swdisp}) it follows that for this model the single-mode-approximation
gives $ 4 \pi \rho_s =  g_3 V_2 $.  The spin stiffness is 
proportional to the probability of finding pairs of electrons 
in the ground state with relative angular momentum $3$ and is 
due to the fact that, in long wavelength spin-wave states, some of these pairs 
make transitions to relative angular momentum 2.  Using\cite{mitra} 
$g_3 \approx 1.5$ we find that the single mode estimate of 
$8 \pi \rho_s$ for the extended hard-core model is $\approx 3 V_2$,
considerably larger than a separate estimate extracted, $\approx V_2$,
from finite-size spin-wave energies\cite{unpub};
apparently the single-mode-approximation
is not quantitatively accurate for this model.  Since, in the large $K$ limit,
the interaction contribution to the 
quasiparticle-quasihole gap is expected to approach $8 \pi \rho_s$,
the decrease of quasielectron interaction energies with $K$ 
in Table 1 should continue to the large $K$ limit.  The ratio of the 
interaction energy contribution to the gap for $K_h=0$ and $K_e=1$ to
$8 \pi \rho_s$ is $\approx 1.2$ compared to values of $1$ and 
$2$ at $\nu =1$ for hard-core and Coulomb models respectively.

We now discuss numerical results for the ideal 2D Coulomb interaction
model in light of the above.  The quasiparticle energies for this model
are listed in Table II.
As expected from the extended hard-core model example and anticipated
originally by Halperin\cite{halphelvetica}, the interaction energy for
the $K=1$ quasielectron state is substantially below that of the $K=0$ 
quasielectron state. The difference found here, $\approx 0.038
e^2/\epsilon \ell$ is consistent with earlier
estimates\cite{spinreversed,bonesteel}. 
Quasihole states with $K \ge 1$ and quasielectron
states with $K \ge 2$ are skyrmion states.
Neglecting finite width and Laudau level mixing effects\cite{spinreversed},
the $K=1$ quasielectron state is energetically
preferred to the $K=0$ quasielectron state for $g \mu_B B < 0.038 
e^2/\epsilon \ell$, an inequality which is satisfied in GaAs for 
fields smaller than $\approx 40 {\rm Tesla}$.  At low temperatures and 
weak Zeeman coupling, it is a good approximation to ignore the 
$K=0$ quasielectron state.  When this is the case, $M(T)$ 
will be similar to that for the extended hard-core model.  If the 
temperature is larger than\cite{sondhi,GM95}
$4 \pi \rho_s \approx 0.012 e^2/\epsilon \ell$
the energetic splitting of the remaining quasiparticle states will
be ineffective and $M(T)$ will be similar to the $\nu = 1$ free-electron
result.  While the Coulomb model cannot simultaneously satisfy the
limits $k_B T \gg 4 \pi \rho_s$ and $k_B T \ll \epsilon_{qe}(K=0)
 - \epsilon_{qe}(K=1)$, the separation of these energy scales is
sufficient to expect some qualitative similarity between $\nu = 1$
free-particle and experimental $M(T)$'s and that is indeed what is
observed.  The simplest approximation for including interaction
corrections at $\nu=1$ is to account for interaction enhancement
of the spin-splitting gap.  The analog of the interaction 
contribution to the $\nu=1$ spin-splitting
gap is the interaction part of the $K=1$ quasielectron,
$K=0$ quasihole gap which is $\approx 0.05 e^2/\epsilon \ell$.

In order to compare with experiment it is necessary to account for
finite-thickness of the electron layers.  The energy scales mentioned 
above will be reduced by approximately a factor of two, depending 
on details of the sample geometry.  The leading low temperature
dependence of $M(T)$ can in principle be dominated either 
by spin-wave excitations or by particle-hole exciations.  
The former contribution is proportional to 
$ \exp (- \Delta_z/ k_B T)$, while the latter 
is proportional to $ \exp (- \Delta_{K_e,K_h})/2 k_B T)$.
The factor of two reduction in the activation energy compared to the 
gap occurs because the particle and hole can
be located anywhere in the sample and directly reflects the 
itinerant character of the electronic system.  
For typical sample parameters, $\Delta_z$ is smaller than the quasiparticle
gap by a factor of 40 or more at $\nu=1$ and only the physics
of long-wavelength spin fluctuations is important in determining $M(T)$ at
low temperatures.  For $\nu = 1/3$, however, 
the situation is completely different, primarily
because the interaction contribution to the $K_e=1,K_h=0$ gap
is much smaller.  Reducing the Coulomb model estimate of this
quantity by a factor of two to account approximately for 
finite thickeness effects, the particle-hole gap 
is reduced to twice the collective gap and we can expect\cite{kasner} 
particle-hole physics dominate even as $T \to 0$. 
This conclusion is consistent with experimental results for
$M(T)$ at $\nu = 1/3$ which is\cite{barrettyale}, in fact, accurately 
fit by the $\nu=1$ independent particle formula with a 
gap approximately twice $\Delta_z$.  

The above considerations suggest the following qualitative explanation
of the experimental $M(T)$ curves at $\nu = 1/3$.  A strongly 
interacting system at $\nu = 1/3$ is similar to a weakly 
interacting system at $\nu = 1$.  As a consequence $M(T)$ 
falls {\it more} rapidly with temperature than it would for 
a non-interacting system at $\nu = 1/3$.  $M(T)$ follows a 
$ \tanh (\Delta / 4 k_B T)$ form with $\Delta \approx 
2 \Delta_z$ because the effective interactions among 
fractionally charged quasiparticles are relatively weak
and because long-wavelength physics does not dominate 
even at low temperatures.  Of course, this picture does 
oversimplify the real situation in several respects,
a point made most tellingly by considering the 
filling factor dependence of $M(T)$ data at $T=0$.  
Our calculations would suggest that, while the lowest energy
quasielectrons could possibly have $K=1$, the lowest
energy quasiholes should certainly have $K=0$.  However, the 
experimental finding is that the number of reversed spins
per quasiparticle in the ground state 
is $\approx 0.05$ for quasiholes and $\approx 0.1$ for 
quasielectrons.  
This finding can be reconciled with our picture of the 
temperature dependence at $\nu =1/3$ by assuming that 
disorder, which is strong for the localized ground state
quasiparticles, favors quasiparticles with different numbers
of reversed spins at different sites and that disorder 
is much less important for the extended quasiparticles and 
quasiholes excited at finite temperatures.
Nevertheless it must be acknowledged that $K=0$ quasielectrons 
are very likely to be present at finite temperatures and that 
disorder probably also has some importance for the $M(T)$
results at $\nu = 1/3$. 

This work was supported by the National Science Foundation
under grant DMR-9714055 and by MEC of Spain under contract
No. PB96-0085. The authors acknowledge
helpful conversations with Steve Girvin, Sean Barrett, Luis Brey
and Carlos Tejedor.

\newpage

\begin{table}
\begin{center}
\begin{tabular}{|c|ccccc|}
        & $N=4$  & $N=5$  & $N=6$  & $N=7$  & $N=8$  \\ \hline
$E_0$   & 0.0000 & 0.0000 & 0.0000 & 0.0000 & 0.0000 \\
$-\epsilon_{qh}(K=0)$ & 0.0000 & 0.0000 & 0.0000 & 0.0000 & 0.0000 \\
$-\epsilon_{qh}(K=1)$ & 0.0000 & 0.0000 & 0.0000 & 0.0000 & 0.0000 \\
$-\epsilon_{qh}(K=2)$ & 0.0000 & 0.0000 & 0.0000 & 0.0000 & 0.0000 \\
$\epsilon_{qe}(K=0)$ & 71.671 & 65.269 & 60.882 & 58.012 &        \\
$\epsilon_{qe}(K=1)$ & 1.4281 & 1.3917 & 1.3610 & 1.3406 &        \\
$\epsilon_{qe}(K=2)$ & 1.3523 & 1.2984 & 1.2621 &        &        \\
$\Delta(K_e=0,K_h=0)$& 71.671 & 65.269 & 60.882 & 58.012 &  \\
$\Delta(K_e=1,K_h=0)$& 1.4281 & 1.3917 & 1.3610 & 1.3406 &  \\
\end{tabular}
\end{center}
\caption[]{
Ground state energy, quasiparticle and quasihole
creation energies with small numbers of reversed spins ($K$), 
and particle-hole excitation energies
for small numbers of particles on a sphere at $\nu = 1/3$. These
results were calculated with $V_0=V_1=100$, $V_2=1$ and all other 
pseudopotentials set equal to zero and do not include
the Zeeman contribution , $K \Delta_z$ for a quasiparticle
with $K$ reversed spins.
The $K=0$ quasielectron energies
will diverge as $V_0$ and $V_1$ approach $\infty$.}
\label{table1}
\end{table}

\begin{table}
\begin{center}
\begin{tabular}{|c|ccccc|}
        & $N=4$  & $N=5$  & $N=6$  & $N=7$  & $N=8$  \\ \hline
$E_0$   & 1.8711 & 2.8056 & 3.8716 & 5.0605 & 6.3626 \\
$-\epsilon_{qh}(K=0)$ & 0.0849 & 0.0978 & 0.1100 & 0.1213 & 0.1320 \\
$-\epsilon_{qh}(K=1)$ & 0.0984 & 0.1101 & 0.1218 & 0.1328 & 0.1433 \\
$-\epsilon_{qh}(K=2)$ & 0.1035 & 0.1144 & 0.1263 &        &        \\
$\epsilon_{qe}(K=0)$ & 0.1968 & 0.2044  & 0.2118 & 0.2213 & 0.2306 \\
$\epsilon_{qe}(K=1)$ & 0.1453 & 0.1573 & 0.1693 & 0.1807 & 0.1917 \\
$\epsilon_{qe}(K=2)$ & 0.1370 & 0.1501 & 0.1623 &        &        \\
$\Delta(K_e=0,K_h=0)$& 0.1118 & 0.1066 & 0.1018 & 0.1000 & 0.0986 \\
$\Delta(K_e=1,K_h=0)$& 0.0604 & 0.0595 & 0.0593 & 0.0594 & 0.0597 \\ 
\end{tabular}
\end{center}
\caption[]{
Ground state energy, quasiparticle and quasihole
creation energies with small numbers of reversed spins ($K$), 
and particle-hole excitation energies
for small numbers of particles on a sphere at $\nu = 1/3$.  Several
natural choices exist for the definition of the quasiparticle energies
which shift $\epsilon_{qe}$ and $-\epsilon_{qh}$ by identical 
constants but give the same values for the quasiparticle-quasihole 
gap $\Delta = \epsilon_{qe} + \epsilon_{qh}$.  The numbers listed 
here are neutral quasiparticle energies in the terminology of 
Morf and Halperin,\onlinecite{qpengdef} but are listed without
the usual neutralizing background contribution.
Without this contribution the ground state energy has 
an electronstatic contribution $\propto N^{3/2}$ and 
the quasiparticle energies have electrostatic contributions
$\propto N^{1/2}$.  These contributions cancel in $\Delta$,
however, which approaches a finite value for $N \to \infty$.
We have not included the Zeeman contribution which, for  
a quasiparticle with $K$ reversed spins, is $K \Delta_z$.}
\label{table2}
\end{table}

\end{document}